# SOFTWARE-DEFINED ELASTIC PROVISIONING OF IoT EDGE COMPUTING VIRTUAL RESOURCES

Patrícia Cardoso
Instituto de Telecomunicações
Lisbon, Portugal
pgcoo@iscte-iul.pt

José Moura
ISCTE-IUL
Instituto de Telecomunicações
Lisbon, Portugal
jose.moura@iscte-iul.pt

Rui Marinheiro
ISCTE-IUL
Instituto de Telecomunicações
rui.marinheiro@iscte-iul.pt

***Abstract*— The fast growth of Internet-connected embedded devices demands for new capabilities at the network edge. These new capabilities are local processing, efficient communications, and resource virtualization. The current work aims to address these capabilities by designing and deploying a new management proposal, which offers on-demand activation of offline Internet of Things (IoT) fog computing assets via a Software Defined Networking (SDN) based solution combined with containerization and sensor virtualization. We propose a testbed as a proof of concept for the main functionalities of this novel solution. The obtained results evidence that the current SDN-based solution can deploy with success activation policies on computational edge containers, which are located within the same network domain of the SDN controller. In addition, different application-level scenarios are also investigated.**

***Keywords— Management; Elastic Provisioning; Software-Defined Networking; Internet of Things; Edge Computing; Container; Self-Activation; Containerization***

## I. INTRODUCTION

The International Data Corporate estimates that at the end of 2025 there will be 41.6 billion of things connected to the Internet [1], generating 79.4 zettabytes of data. These predictions will be supported by two evident pillars: i) the improvements in the Telecom sector, with more ubiquitous and cheaper Internet access; and ii) the edge-cut technologies to produce smaller and more powerful processing devices.

The enormous scale of connected devices, the management of very large amounts of generated data, the use of different communication protocols, hardware and software supplied by distinct manufacturers, makes ineffective the operation of legacy network architectures. As such, there is a requirement for a more flexible and comprising network architecture, essentially at the network edge.

One of the most promising approaches to address the above-mentioned challenges is the addition of an edge layer between IoT devices and the cloud to support processing, communication, and storage services near the end user. This new layer can increase the performance, mobility, security and privacy in the IoT, as well as reduce the data volume exchanged through the backhaul links and guarantee a low latency to the data access [2]. However, the addition of this edge layer in the network infrastructure also introduces new challenges to the management of the available resources at that peripheral infrastructure [3][4] and associated services, which could be offered in a federated way [5].

The available edge computational resources, sometimes also referred as fog computing (FC), should be managed in a distinct way of the computing assets from remote cloud data centers, because edge servers have fewer available processing resources when compared to cloud servers [3]. In the current paper, we use FC and Edge computing (EC) concepts in a similar way to model scenarios with computational containers operating at network peripheral domains. Nevertheless, the reader is referred to [6] that discusses FC and EC as slightly distinct but relevant pillars of the current evolution on networked systems to satisfy the major requisites imposed by emerging use cases, e.g. involving IoT devices.

Software-defined networking (SDN) has emerged as a network architecture that addresses the complexity on the management in FC environments, trying to solve the IoT heterogeneity problem. SDN enables the creation of independent features and innovative open-source services in a very easy way, overcoming the network ossification imposed by the legacy proprietary network services [7]. In traditional networks, the control plane and the data plane are located within the network devices, requiring a personalized configuration on each device, using a low-level and often vendor-specific commands. SDN takes out the control plane from the data plane devices. In this way, it is offered a centralized logical control and some abstraction from the complex functional aspects on the data plane. Thus, an SDN-based solution can optimize traffic management as well as logically support services requirements from a centralized user interface (UI). It also offers a high level of agility, traffic programmability and the more recent opportunity for implementing network automation or network programmability [8].

Additionally, the use of virtualization techniques applied to SDN and sensor services can tackle the scalability and heterogeneity issues enforced by most upcoming IoT-based scenarios [9]. The use of virtualized sensors provides hardware and software abstraction, reduces the number of physical devices needed and allows their resource management, such as through open APIs [10][11].

The current paper addresses the IoT and sensor networks challenges by proposing not only the design, but also an implementation of an approach, which performs on-demand activation of offline IoT edge computing containers. This is achieved by using an SDN controller pushed to the edge that is orchestrated by a NorthBound Broker entity. The Broker manages the lifecycle of edge IoT services that run within containers. To save energy, a service container is initially stopped, but it can be activated by the Broker, in reaction to clients requesting for that service. This service is embedded in edge devices or middleboxes, such as wireless access routers or set top boxes at user premises, using a lightweight full operating system virtualization environment. In the current work, we also discuss performance and functional results obtained from evaluation tests made on the proposed software-defined solution augmented by the Broker. We highlight some positive outcomes from applying this solution in future IoT scenarios with embedded devices at the network edge, offering and/or using innovative services/applications.

We have obtained several pertinent and novel outcomes, which are divided in two parts such as, system performance and system functionality. Considering the system performance, our investigation has provided the next evidences: i) even when network data processing and virtualization resource management are pushed to the edge, proactive control modes have significant performance improvement over others, such as reactive or hybrid; ii) the time to activate computing nodes at the network edge as well as the time to activate network edge nodes and then these establish their control channel with the SDN controller, should both be considered in the activation delay of data layer nodes (and/or services) that are initially powered off; and iii) the elastic management of edge resources can bring significant advantages in energy efficiency, e.g. high autonomy of battery-operated devices, low cost of system operation / maintenance.

The novel outcomes of our work are as follows: i) it presents an accurate description of a SDN testbed where a containerization technology has been used to provide microservices to the edge of the network; ii) the proposed solution uses a novel SDN-based mechanism to orchestrate computational instances deployed at the network edge; iii) the testbed results evidence the current SDN-based solution is able to deploy with success activation policies on computational edge containers, which are located within the same network domain of the SDN controller.

The paper structure is as follows. Section II revises the more relevant related literature, highlighting the novelty of our work. Section III presents the design and implementation details about the proposed solution. The evaluation results of this proposal are discussed in Section IV. Finally, Section V concludes and points out some future research guidelines.

## II. RELATED WORK

This section discusses relevant related work, highlighting the main novel aspects of the present research. From the literature, there are several emerging paradigms that can be used as enablers for the next generation of wireless sensor networks. In particular, the beneath text selects data-driven softwarized solutions relevant for IoT-based scenarios to ensure various goals such as control, orchestrate, or abstract the available network edge assets, mainly computing ones.

The work in [12] discusses adaptable and data-driven decision making for communication systems. They propose Machine Learning (ML) [13] modules to enhance the functional primitives of observation (ensured mainly via SDN), composition (ensured mainly via NFV), and control (ensured by the coordination between SDN and NFV) in the presence of uncertainty in relation to the network status evolution. Offering the previous enhancements, the data-driven networked systems can learn and properly react to changes on the networking context as well as to unexpected variations on traffic load. In addition, [14] proposes a smart SDN Management of Fog Services. The work in [15] studies the orchestration of SDN applications that cooperate to offer network-wide resilience. The studied applications involve traffic classification, anomaly detection, and traffic shaping.

Kobo et al. [10] studies SDN as a technology enabler for the upcoming use cases of wireless sensor networks. The work in [16] investigates SDN for IoT applications. They combine SDN with virtualization frameworks, such as NFV and network slicing. NFV can use high-level policies to manage IoT resources and network slicing can support QoS/QoE at the network edge. The authors of [17] study the security benefits of the cooperative operation of SDN and NFV to enable the efficient operation of IoT networks.

TABLE I. LITERATURE COMPARISON (+ MEANS COVERED)

| Reference | SDN | NFV (Broker) | ML | Orchestration | Uncertainty | Offloading | Network Slicing | Docker | LXC | On-demand activation of IoT Services |
|---|---|---|---|---|---|---|---|---|---|---|
| [10] | + | | + | | | | | | | |
| [12] | + | + | + | | + | | | | | |
| [13] | + | | + | | | | | | | |
| [14] | + | | + | | | | | | | |
| [15] | + | + | | + | | | | | | |
| [16] | + | + | | | | | + | | | |
| [17] | + | + | | | | | | | | |
| [18] | + | | | | | + | | | | |
| [19] | + | | | | + | | | + | | + |
| [20] | + | | | + | | | | | | |
| [21] | | | | | | | | | + | |
| [22] | | | | | | | | | + | |
| Our work | + | + | | + | + | | | | + | + |

Tomovic et al. [18] have designed a solution that combines the major benefits normally offered by both SDN and FC. Their proposal orchestrates fog resources, via SDN controllers, to diminish the level of complexity to efficiently control those resources. In addition, the SDN scalability issue is tamed by delegating some controller's processing tasks to fog computing nodes. The virtualization is out of scope of their work.

Another approach involving SDN and container-based technology is the one described by Xu et al. [19] which proposes an in-house controller for elastically managing Docker computing resources at edge switches. They have used an SDN controller pushed to the edge, to manage the life cycle of services. Our research is also aligned with the main objective discussed in [20], which adopts localized flow management performed by distributed controllers to overcome the additional delay imposed by the constraints on the control channel between each SDN controller and the SDN-based switch. However, Xu proposal was supported in the Docker platform. By far, Docker is the most popular implementation for light virtualization, but focused on a one-app-per-container deployment. This makes it more difficult to group and merge heterogeneous data from several IoT services towards the synthetization of sensor-based information. In addition, a Docker container runs at the user-space, which increases both the system overhead and the activation time of that container. Clearly, these functional aspects need to be enhanced.

Linux Containers (LXC) on the contrary of the Docker case are intended to be mainly used as full-fledged machines, rather than running only one application, and have much lower performance overhead. It has been proven that LXC is the most performant solution in nearly all virtualization scenarios [21]. Our work consubstantiates the importance of more performant containers to be used at the edge in accordance to the previous results available in [22]. They investigate the deployment of LXC virtualization as a more light and agile container-based solution than the one offered by Docker. However, they have not considered SDN. Table I compares the current work with previous one, highlighting the novel aspects offered by our contribution.

Recent work as [7][23] provide comprehensive literature discussion in how emerging IoT services can be enhanced through the collaborative deployment of both SDN and edge computing. For providing a solution that solves the different challenges present in IoT environments, the current work proposes a novel integration of SDN with on-demand activation and orchestration of light containers, in which some of these can be virtualized sensors. The design and deployment of our proposal are discussed in the next section.

## III. PROPOSED SOLUTION

This section has two parts: i) the first part debates the design; ii) the second part discusses proposal deployment.

### A. *Architecture*

Figure 1 presents the high-level design of our proposal. It is a FC system that combines the SDN paradigm with virtualized processing resources to be managed in IoT-based use cases. The novelty of this SDN-based proposal is how it extends the control of networking assets, which is typically the focus of most of the previous work in the literature, to support the management of edge-computing resources in elastic and agile ways. A possible distributed administration of edge storage is possible, eventually in reaction to the evolution of spatio-temporal data popularity [24]. Popularity is out of scope of the present work, but we subscribe that a tight coupling of a SDN controller and a Broker in edge computing nodes, such as middleboxes at user premises, is key for on demand management of local resources, such as in [19].

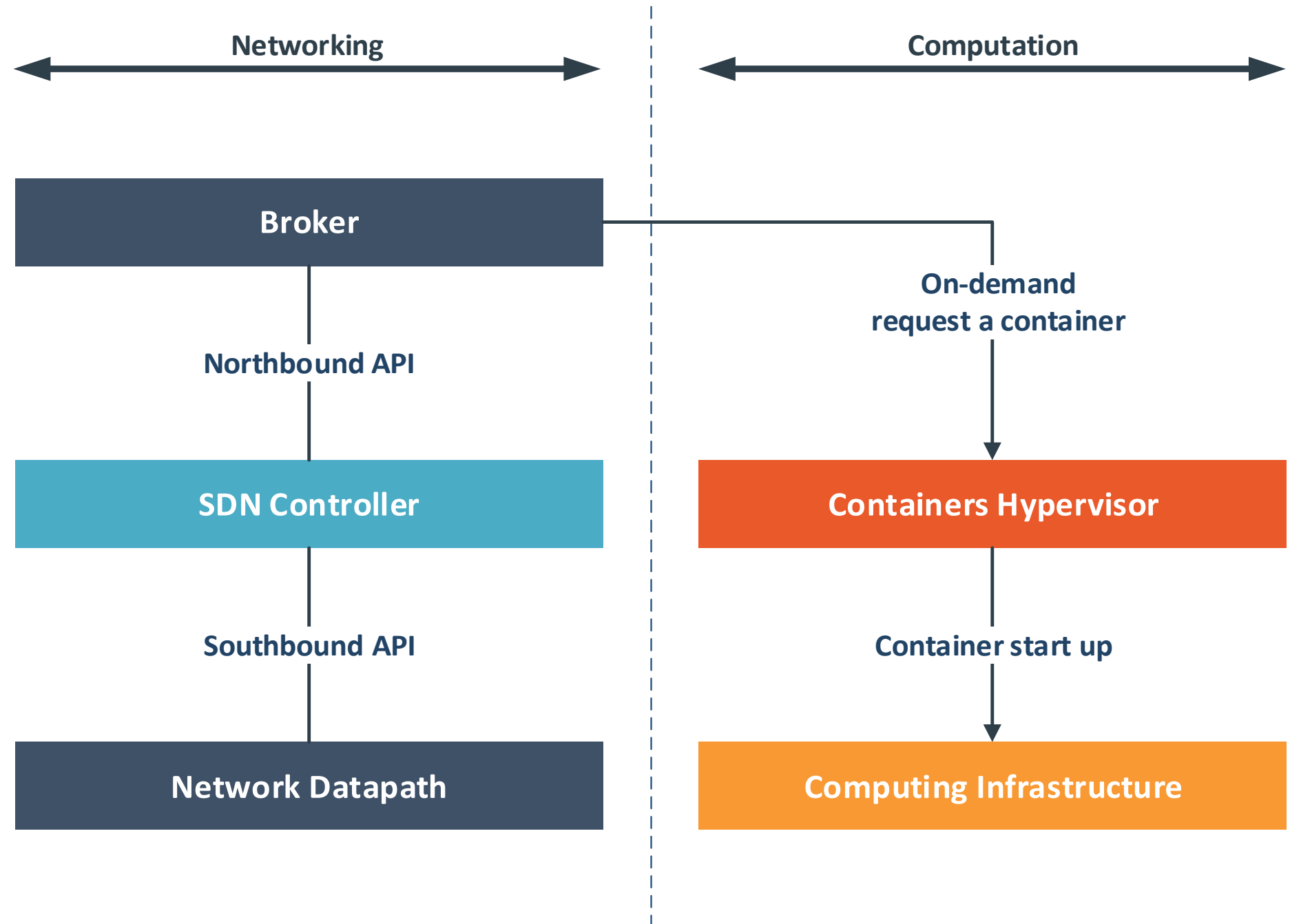

Fig.1. Proposed Architecture

In our proposal, the SDN data plane is formed by switches, which communicate with an SDN Controller through the Southbound API. In the application plane there is a Broker that analyzes previously selected data plane messages and, as necessary, the Broker boots containers to process those messages. The Broker communicates with the controller through the Northbound API. In the beginning, not all containers at the data plane are running, enhancing the system sustainability and the energy efficiency in a similar way to what has been proposed before for data center networks [25].

The Broker entity that orchestrates with the SDN Controller can be a self-driving agent, which deals with the uncertainty at the network datapath status evolution, as shown in Figure 2. This datapath status uncertainty is due to [12]: arrival of new flows, flows traffic, and container availability. The broker behavior regarding the network datapath is represented as an observation-action loop with some naïve self-adaptive characteristics, to deal with the system uncertainty. The SDN Controller is not in Figure 2 for simplifying its visualization.

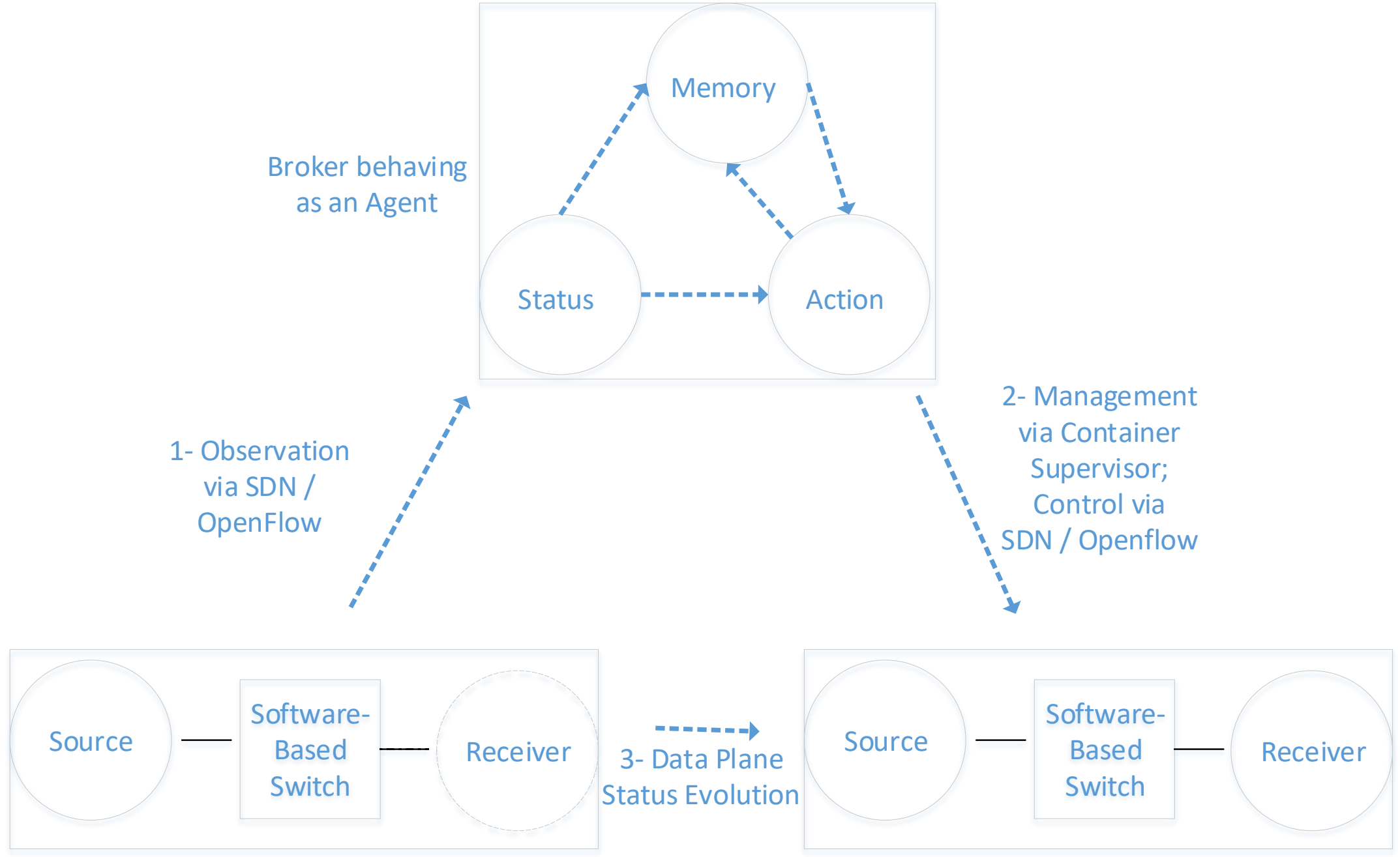

Fig.2. Broker Behaving as an Agent [12]

Analysing Figure 2, one can conclude that the higher-level entities of our system, which are the Broker and the SDN Controller, receive status information from the context around network devices. This information is obtained during the functional phase designated as Observation (phase 1 of Figure 2). Then, the Broker entity, according to the historical information about the previous actions made on the datapath plane, combined with either the current statistics from the resource usage or the signaling of new data flows, it decides to (or not) send new control messages to the network datapath (phase 2 of Figure 2). In the case new control messages are sent to the network datapath, its status will naturally evolve, which is the situation represented as phase 3 in Figure 2, when the Receiver container is activated.

The observation-action loop can be analyzed as a probabilistic channel [12]. In this context, the capacity of that channel is given by the empowerment variable [26]. The empowerment is the measure of how much influence the Broker has on its controlled environment given the diverse characteristics of the Broker, such as, the Status, Memory, and Action. The main goal of the Broker is to select an action impacting in the data plane status with the highest empowerment. Nevertheless, this aspect is out of scope of the current work. Further details on this are available in [12].

The current work empowers the Broker to elastically boot IoT containers at the data plane. A similar solution, with minor adaptations, can also be applied to scale out at the control plane, by activating SDN controllers running within containers, following greater demand increases for new flows at the data plane. Nevertheless, the on-demand activation of virtual entities at the control plane is not covered by the current paper. Further details are in [27].

### B. *Deployment*

Figure 3 details the deployed architecture of the current proposal at the network edge. This implementation of an edge computing node, such as an edge middlebox, is composed by an SDN controller, a Broker, a virtual multilayer switch, and service containers.

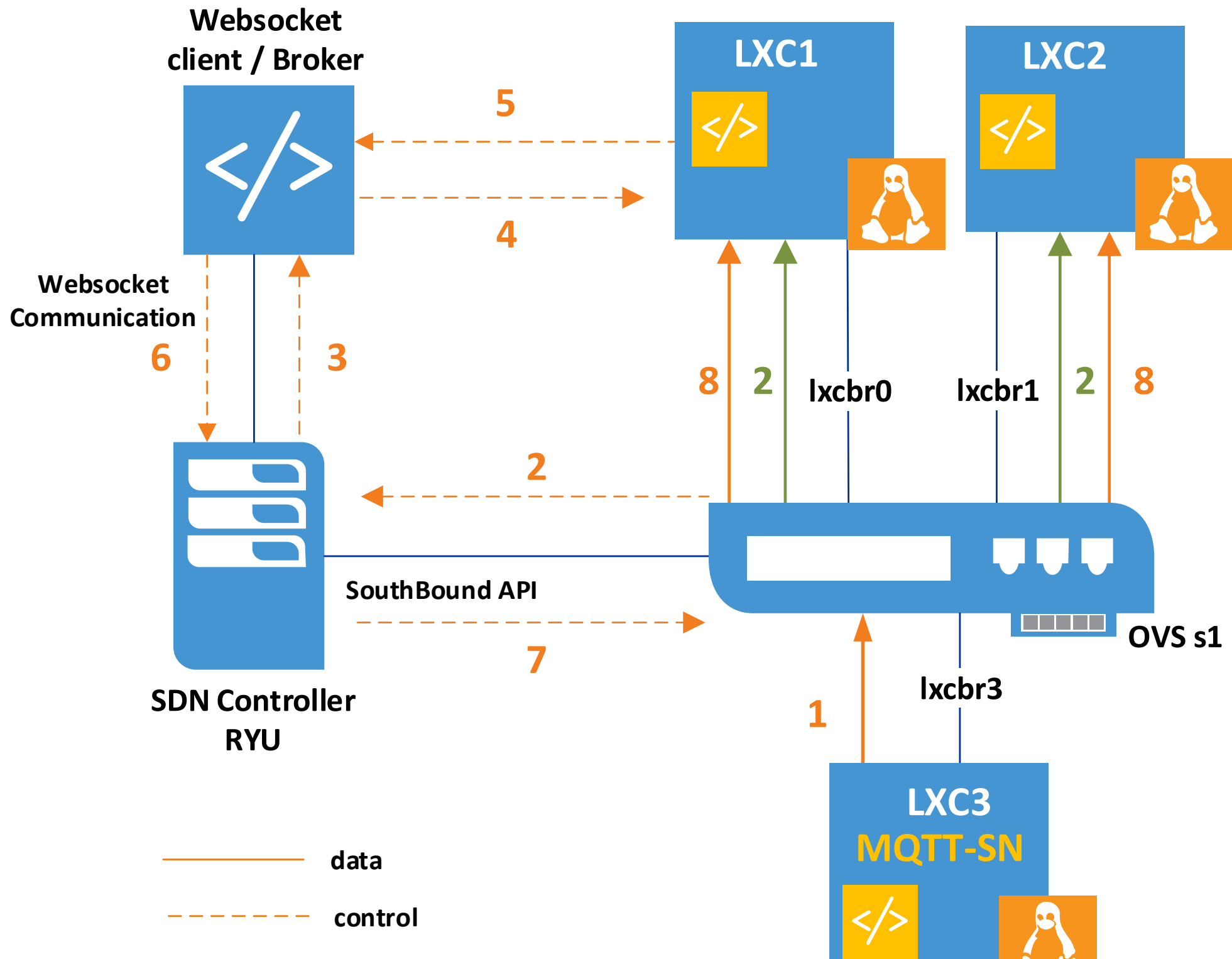

Fig.3. Detailed Deployed Architecture

This implementation materializes a testbed to obtain performance and functional results from the evaluation of our proposal. It uses a Ryu SDN controller[1]. This is an open-source controller, that is lightweight and agile, key for deployment at the edge, where resources are more limited and heterogeneous. Three Python scripts were coded to study three distinct controller behaviors, i.e., reactive, hybrid and proactive mode. Regarding containerization, Linux containers were chosen as they are a more lightweight and agile option than other alternatives, such as the Docker. The performance advantages of using Linux containers occur because they run directly inside the Kernel. The containers LXC1 and LXC2 virtualize IoT sensors and, as already mentioned, these containers are not always running. The virtual multilayer switch is an Open vSwitch[2] (OVS), controlled by the Ryu SDN controller. The SDN controller uses Websocket NorthBound communication with the Broker, that is responsible for the management of the container's lifecycle. All previously mentioned components are installed within our middlebox, a small form factor single-board computer running Linux.

In more detail, the network topology (i.e. architecture data plane) includes one Open vSwitch (OVS s1), one namespace host (h1) and diverse links/bridges, all configured through shell scripts with network namespaces, bridge control and Open vSwitch commands. The host h1 was created with the intent to test Ryu scripts, mainly the switch flow entries, through ping and iperf commands. Even though containers were previously created, it was necessary to connect them to the Open vSwitch through Linux bridges. The OVS s1 was configured with a static datapath-id value of one, the OpenFlow protocol version 1.3 and a logical transport connection to the Ryu controller in the TCP port 6633, which is the default port for OpenFlow from the SDN controller side. Further details about this communication will be given in the next sub-section.

For our implementation, a use case scenario has been used, where external clients, in this case LXC3, request data from virtual sensors, either LXC1 or LXC2. To ensure the communication between sensors and clients we have adopted the MQTT-SN[3] protocol. We have made this choice, because it is the most suitable application-layer protocol in scenarios where a permanent connection is not possible or desired. This is valid because this protocol, created for sensor networks, uses UDP as the transport protocol. In this way, MQTT-SN avoids establishing logical transport connections before data exchange, and it demands less system resources for the transport layer operation. The former characteristic diminishes the data transfer delay, and the latter one offers a more scalable solution.

[1] https://ryu-sdn.org
[2] https://www.openvswitch.org

In our testbed scenario, the containers LXC1 and LXC2 virtualize IoT sensors that are activated on demand and, as already mentioned, in the beginning they are not running. The LXC3 is the container which has installed the MQTT-SN Broker. This container is always running. It behaves like an intermediary node that requests data from sensors. When triggered by an IoT consuming client (not shown in Figure 3 for simplicity purposes), the LXC3 sends a MQTT-SN message to either LXC1 or LXC2 sensor (step 1). When the OVS s1 receives the request packet, it will check if there is a forwarding rule in the OpenFlow table. If it finds a match entry, it will send the packet to the destination container (step 2). However, if there is no match in the flow table, the switch encapsulates a copy of that packet within a packet-in message, which is sent to the controller, querying which action should be taken (step 2). Through a Websocket connection, the Ryu controller forwards the message to the Broker (step 3). In this way, the Broker can analyze the received message and activate the target container (step 4). Once the Broker receives the feedback, notifying that the container is running (step 5), the Broker informs the controller (step 6). At this step, the controller sends a flow rule to be stored in the switch as well as the forwarding action to be applied to the received packet. Both messages are sent to the OVS s1 (step 7). At last (step 8), switch s1 sends the request to the target sensor container.

*Switch and SDN Controller Communication*

As described above, three different Ryu (Python) scripts were deployed. In the reactive mode script, when the Ryu controller is started, a flow rule (i.e. flow miss default rule) is installed into the switch. In this way, the switch sends a packet-in message to the controller whenever the switch (OVS) receives a packet. In reaction to each received packet-in, the Ryu script uses the class OFPActionOutput and the flag OFPP_FLOOD in the packet-out message, specifying to the switch that the received packet needs to be forwarded to all the available switch ports except the one used to receive that packet.

[3] http://mqtt.org/documentation

In the hybrid script, the system behaves, at first, in a similar way when compared to the reactive mode. The flow miss default rule is the only rule installed into the switch when the Ryu controller is started. However, in the hybrid mode, the reactive behavior only happens for the first packet of a specific flow. In fact, after the first packet-out message, the Ryu controller sends a flow-mod message with the out_port flag enabled. In this way, the Ryu controller indicates to the OVS through which port the next packets of the same flow should be forwarded. This new action belongs to a flow rule, which is installed into the switch flow table, avoiding future flooding and allowing a faster and more efficient transmission of the packets of the same flow by the switch, because the forwarding decision is made locally, not involving the SDN controller.

The proactive script was written to install into the switch not only the "ask to the controller" rule, but also the rules that indicate the port where the flow should be forwarded to reach pre-defined Linux containers. By using this script, it is expected a lower number of packet-in / packet-out messages between the OVS and the controller, allowing even better results when compared to the hybrid mode. The next sub-section gives some information how the Ryu controller and the Broker communicate between them.

*Communication between the SDN Controller and Broker*

Whenever the SDN controller receives a packet from the switch, the controller sends a copy of the header packet to the Broker through a Websocket connection. The Broker then analyses the header fields and extracts useful information about namely the physical and IP addresses, the transport protocol and its ports. Figure 4 shows an example of a message transferred from the controller to the Broker.

Each Linux container has known IP and physical addresses. When an ARP message is identified, and the destination IP address matches one manageable LXC, that container is started by the Broker. If the container is already running or if the destination IP does not match any of the pre-defined containers, the system logs corresponding messages for future analysis. This allows, for instance, the registration of the communication activity of individual virtual sensors to deactivate dormant containers. The next sub-section debates the communication among IoT devices.

```
{"params": ["{\"msg\": \"0xff 0xff 0xff 0xff 0xff 0xff 0x00 0x00 0x00
0x00 0x00 0x26 0x08 0x06 0x00 0x01 0x08 0x00 0x06 0x04 0x00 0x01 0x00
0x00 0x00 0x00 0x00 0x26 0x0a 0x00 0x03 0x1a 0x00 0x00 0x00 0x00 0x00
0x00 0x0a 0x00 0x03 0x15\"}"], "jsonrpc": "2.0", "method": "send", "id
": 1}
ARP Message
src IP: 10.0.3.26
dst IP: 10.0.3.21
STARTING TEMPERATURE SENSOR
```

Fig.4. Exchanged Message from the controller to the Broker

*MQTT-SN communication with IoT virtual devices*

Figure 5 exemplifies how IoT containers exchange messages, where LXC3 publishes messages under the topic “askTemp” and keeps subscribing the topic “sendTemp” until it receives a message under that topic. When the LXC1 sensor receives a message requesting the temperature, then LXC1 publishes the temperature value in the correct topic.

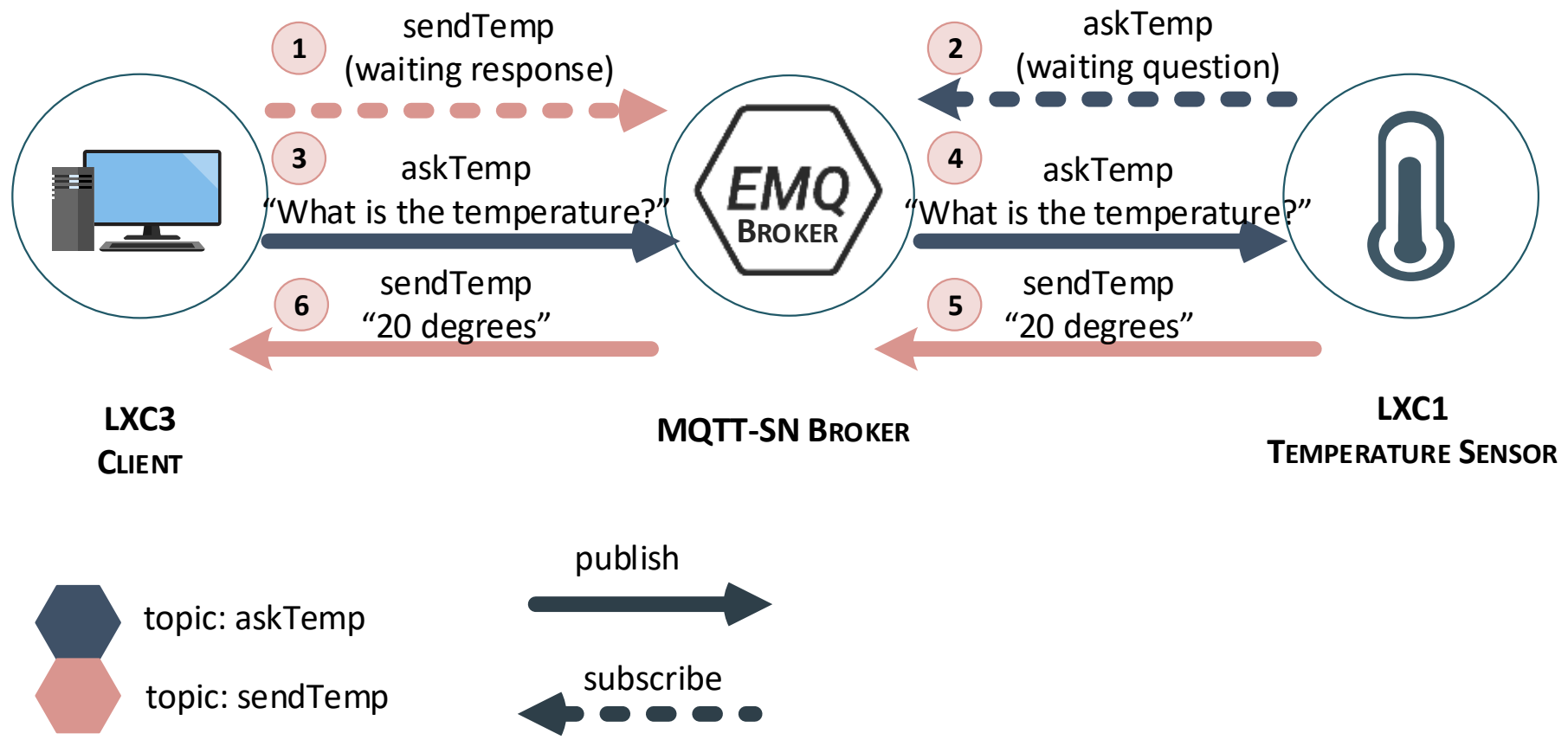

Fig.5. MQTT-SN sensor/client communication scenario

The exchange of MQTT-SN messages between IoT containers requires the installation of a message broker and MQTT-SN clients. The EMQ4 broker was selected due to its versatility supporting several IoT protocols, such as MQTT, MQTT-SN and CoAP. It is an open source IoT MQTT message broker based on Erlang/OTP platform. To access the features of MQTT-SN, the use of the EMQ-SN plugin is required. EMQ recommends several MQTT-SN clients, from which the one from MQTT-SN tools was chosen.

[4] https://www.emqx.io

These tools support some MQTT-SN features, e.g. as QoS -1, 0 and 1, publishing retained messages, short topic IDs, amongst others, but having the drawback of not allowing the QoS 2.

## IV. RESULTS AND DISCUSSION

This section presents how performance and functional tests were made over the implemented proposal and discusses the more relevant results that were obtained. Figure 6 further details the initial testbed, which consists of a network topology formed by a software-based switch (OVS s1), a host (h1), and several containers (LXC1, LXC3). In addition, the resources of the network topology are managed by high-level entities such as Ryu SDN Controller and a Broker. The sub-section A discusses performance results regarding network configuration and container activation. The sub-section B analyses how our proposal behaves in a more realistic scenario, submitting our system to more stressful tests. The sub-section C discusses functional results obtained when an IoT protocol is used between the sensors and clients. Finally, the sub-section D compares the performance of several congestion algorithms in a testing scenario with constrains in both rate and latency.

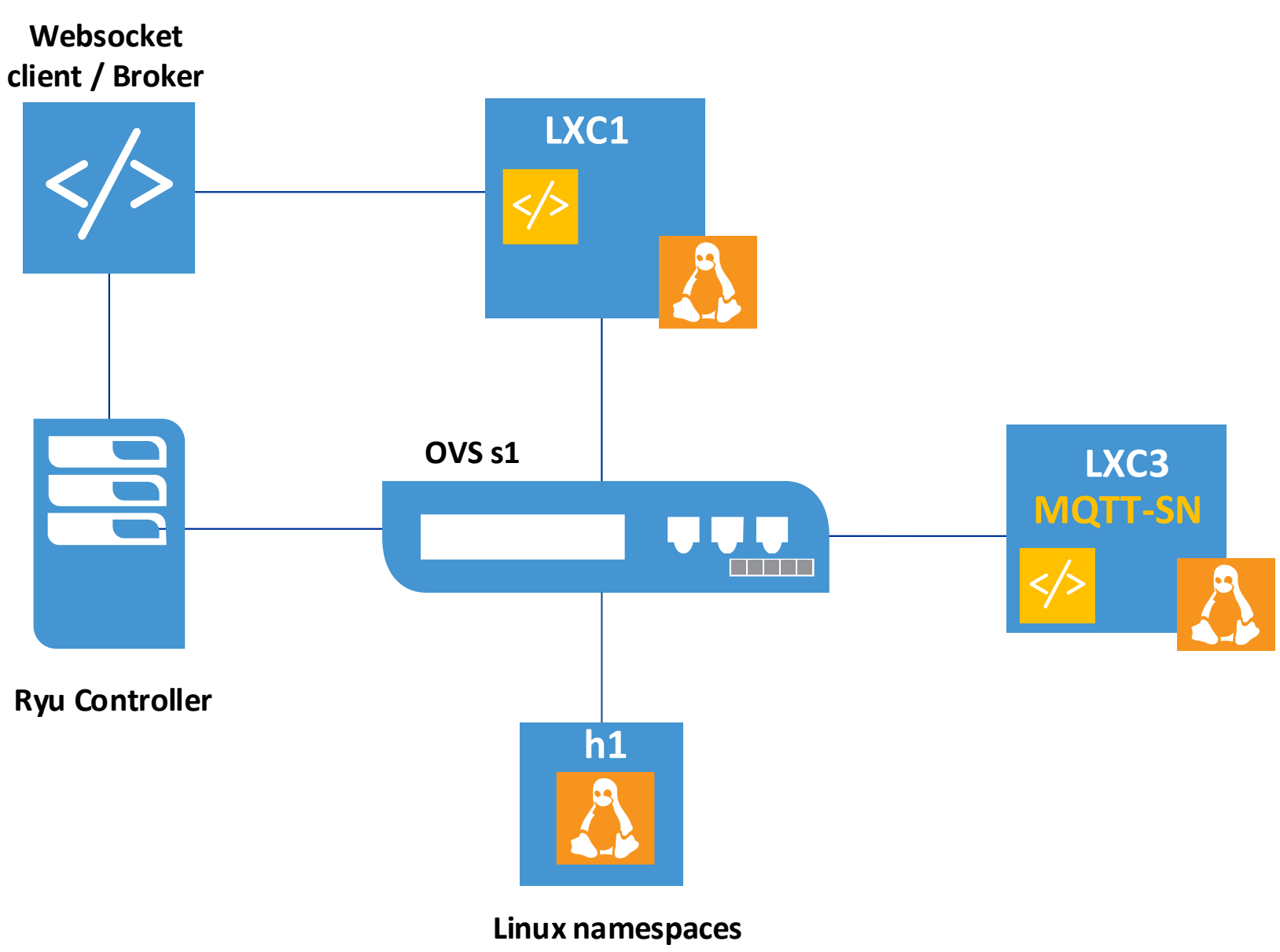

Fig.6. Initial testbed

### A. *Delay Performance Tests*

The current sub-section presents and discusses evaluation results to assess the impact of the SDN controller's behaviour on the time required to process data plane messages and the time it takes for the Broker to start up containers, which deploy virtual sensors, and send an event to the SDN controller to proceed with the communication through the data plane. This is particularly relevant, to demonstrate the feasibility of on-demand management of resources that run on constrained devices at the network edge, such as small form factor single-board computers.

We have tested three distinct SDN controller´s operation modes, namely reactive, hybrid and proactive. In each operation mode, delay communication results were obtained in two scenarios: i) communication between a host device (i.e. using Linux network namespace) and a sensor container; and ii) communication between an external client and a sensor container.

The results for each scenario were obtained by sending four ICMP packets. In the first scenario, the host h1 sends ICMP packets to the sensor's container (LXC1) whereas in the second scenario this communication is started by the client at LXC3. The presented results are averages of the samples obtained via 50 runs of the same testing scenario. The next results are about the reactive control mode.

*Reactive Mode*

In the reactive mode, only the default flow miss rule is installed by the SDN controller in the switch. Here, it is expected that a considerable number of control messages are exchanged, through the control channel, between the SDN controller and the software-based switch.

As an example, in the capture depicted in Figure 7, the RTT associated with the first ARP message (first packet-in and packet-out) was 506ms, whereas the RTT of the second ARP message was 109ms. The overtime expressed by the first ARP message is associated with the activation time of the Linux container, which falls in the range [150, 400] ms.

| Time | Source | Destination | Protocol | Info |
| --- | --- | --- | --- | --- |
| *REF* | 127.0.0.1 | 127.0.0.1 | OpenFlow | Type: OFPT_PACKET_IN |
| 0.505992153 | 127.0.0.1 | 127.0.0.1 | OpenFlow | Type: OFPT_PACKET_OUT |
| *REF* | 127.0.0.1 | 127.0.0.1 | OpenFlow | Type: OFPT_PACKET_IN |
| 0.108769250 | 127.0.0.1 | 127.0.0.1 | OpenFlow | Type: OFPT_PACKET_OUT |

Fig.7. Packet-In and packet-Out ARP Messages (Scenario 2)

Figure 8 shows the network path traversed by a pair of ICMP messages (i.e. request / reply). In this mode, every time a new packet arrives at the switch, a packet-in message is sent back to the controller, which in turn responds with a packet-out message directing the switch to flood out the packet. Thus, for every ICMP packet sent, a respective packet-in and packet-out is generated, which is the reason why packets, after the first one, have a non-negligible RTT.

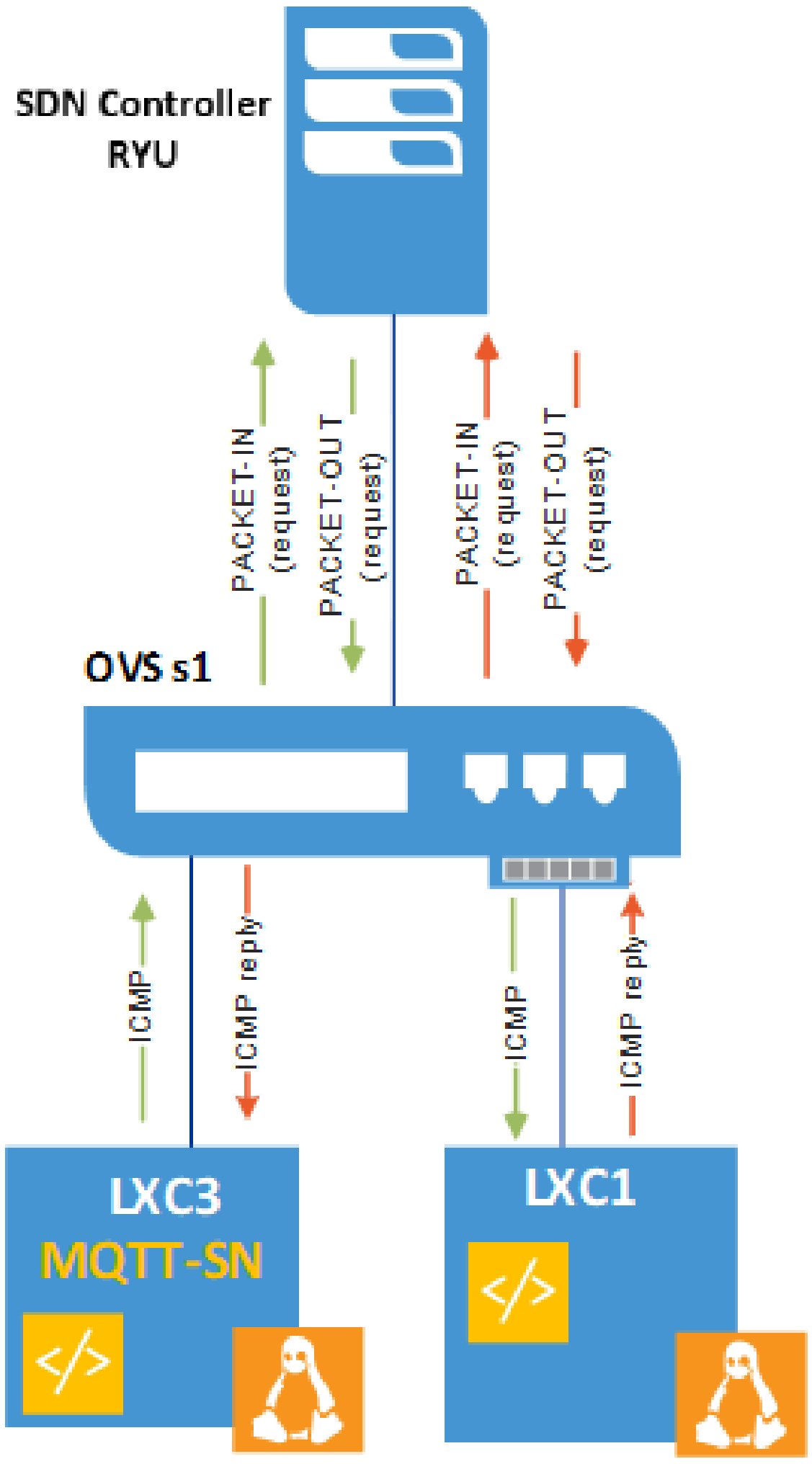

Fig.8. ICMP traffic in the reactive mode (Scenario 2)

Analyzing further results of the reactive mode, presented in Figure 9, we can see the results of both two scenarios which were already explained (i.e. Scenarios 1 and 2). Comparing the results of both scenarios, one can conclude that the first scenario has better response times than the second one. This difference is justified by the fact that in the first scenario the network namespace h1 is a virtualized entity completely embedded within the Linux kernel, in opposition to what occurs in the second scenario, where the container LXC3 has its management software (i.e. Docker) running more slowly in the user space. As it was already explained before (see Figure 7) it is perfectly noticeable that the average response time associated to the first packet, for both scenarios, is higher than the response time associated to remaining packets of the same data flow. In this way, analyzing Figure 9, we can conclude that the first ICMP packet has a higher Round Trip Time (RTT) than the following three ICMP messages because the first ICMP request message can reach its destination (i.e. LXC3) only after the ARP tables have been configured as well as the destination container LXC3 has been launched. The next results are about the hybrid mode of the SDN controller.

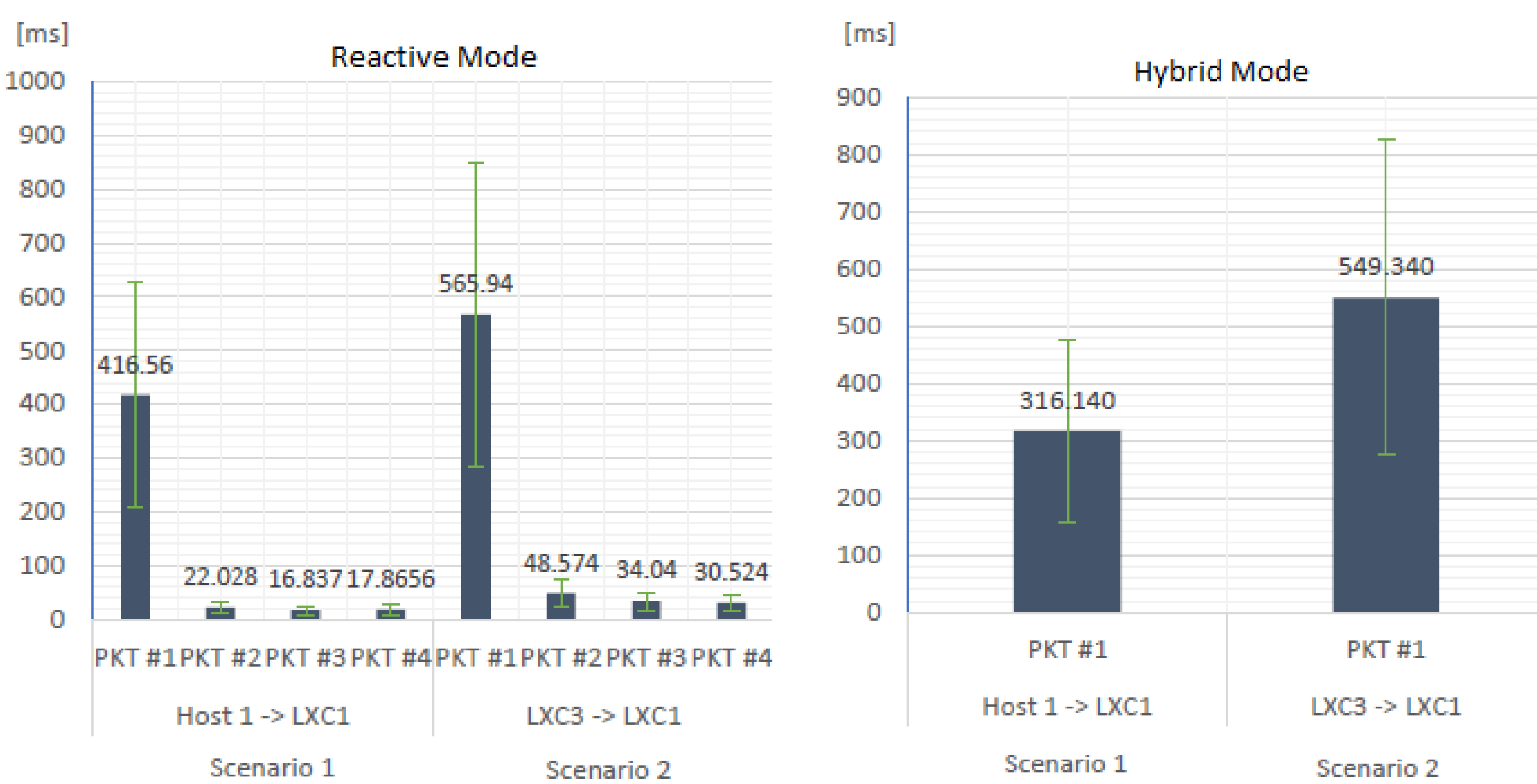

Fig.9. Reactive Mode – Scenarios 1 and 2

Fig.10. Hybrid Mode - Scenarios 1 and 2, 1st packet

*Hybrid Mode*

As mentioned in the previous section, when the Ryu-hybrid mode script is executed, a rule is installed in the switch, with the action to send a packet-in message to the controller if there is no match in the switch flow table. As expected, Figure 10 visualizes that the response time of the first ICMP packet is like the one obtained in the reactive mode, as in both modes the switch in the beginning of the test only has the default rule stored in its flow table.

However, further analysing the results of the hybrid mode in Figure 11, the second and next ICMP packets have a much lower response time, when compared to the equivalent ones of the reactive mode (Figure 9).

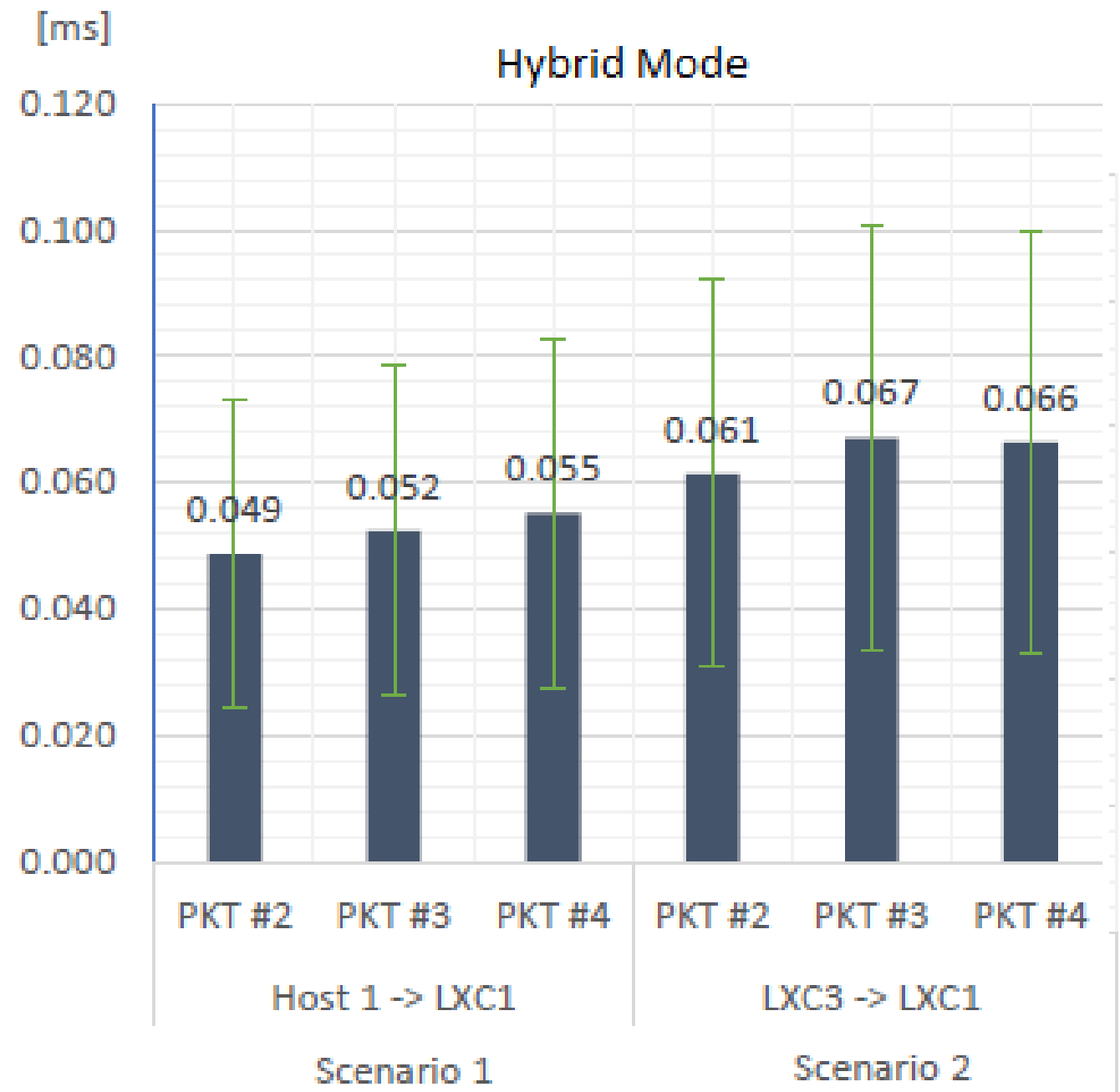

Fig. 11. Hybrid Mode - Scenarios 1 and 2, 2nd to 4th packet

The last referred RTT decrease is due to the on-demand installation of a rule, by the controller, into the switch with the indication of the port from where the traffic should be forwarded. Thus, for all ICMP packets after the first one, the switch does not send back a control message to the controller, because the switch already knows the forwarding port. In this way, the switch immediately sends the packet out through the switch port where the destination container is connected. The next results are about the proactive mode of the SDN controller.

*Proactive Mode*

For the proactive mode, a bootstrap function was added to add flow rules beforehand into the switch flow table. Thus, when the Open vSwitch is instantiated and becomes logically connected to the Ryu controller, those rules are immediately pre-installed on the switch. These rules specify the outgoing ports through where the associated messages should be forwarded, towards the appropriate destination container. This added proactive function is important to avoid unnecessary messages being diverted to the controller via the logical control channel. However, ARP messages rules could not be installed in the switch flow table because the ARP messages are used by the Broker to activate destination containers. The proactive installation of ARP rules in the switch would prevent the ARP message to reach the Broker and impairing the containers automatic boot.

Analysing Figure 12, it is noticeable a relevant improvement for the result of scenario 1 in comparison with the similar result in Figure 10. In fact, the response time of the 1st packet decreased from approximately 300ms (Figure 10) to 200ms (Figure 12). This improvement on the system performance between the proactive and hybrid modes occurred because the initial installation of forwarding rules in the switch increases the number of local forwarding decisions in the switch, optimizing system performance.

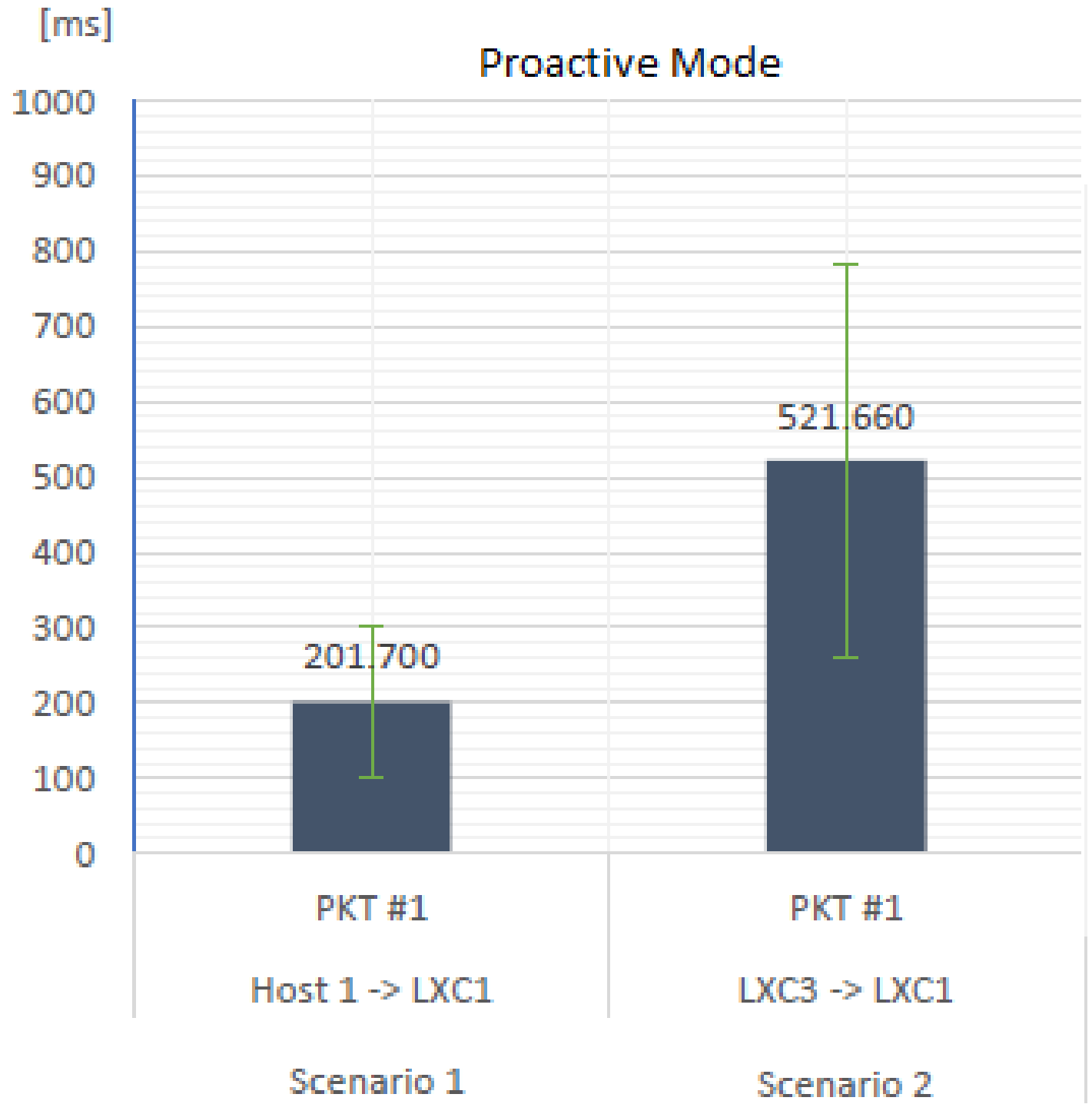

Fig.12. Proactive Mode - Scenarios 1 and 2, 1st packet

*Comparison Among the Three Controller Modes*

Table II compares the performance results of the next controller modes: reactive, hybrid and proactive. Considering the first data request (i.e. first packet), it is noticeable that the proactive mode obtained the best time results. This is due to the pre-installed flow rules in the Open vSwitch, which enables autonomous forwarding decisions at the switch, avoiding the additional communication through the control channel. Regarding the second, third and fourth requests of the same data flow, both proactive and hybrid modes obtained similar results, and much better than in the case of reactive mode. This better performance of both proactive and hybrid modes in relation to the reactive mode is justified again by the former modes can forward a significant number of packets directly to the receiving container, without involving the SDN controller.

TABLE II. PERFORMANCE TESTS COMPARISON (MS)

| Controller's Behavior | PKT | Scenario 1 | Scenario 2 |
|---|---|---|---|
| | | *Host 1 -> LXC1* | *LXC3 -> LXC1* |
| Reactive | 1 | **416,560** | **565,940** |
| | 2 | 22,028 | 48,574 |
| | 3 | 16,837 | 34,040 |
| | 4 | 17,866 | 30,524 |
| Hybrid | 1 | **316,140** | **549,340** |
| | 2 | 0,049 | 0,061 |
| | 3 | 0,052 | 0,067 |
| | 4 | 0,055 | 0,066 |
| Proactive | 1 | **201,700** | **521,660** |
| | 2 | 0,042 | 0,062 |
| | 3 | 0,042 | 0,074 |
| | 4 | 0,041 | 0,064 |

Overall, it is possible to see that the worst response time results were obtained with the reactive mode. It is a simpler and less complex control mode, but by storing in the switch flow table only the default miss-flow rule that sends every received packet to the controller, it implies a high overhead on the number of control messages through the logical channel between the software-based switch and the SDN controller. The reactive mode also implies an increase on the response time of the SDN-based system to perform the forwarding decisions on the data plane to correctly deliver the messages to the destination containers. This communication overhead and the extra processing in both the SDN Controller and Broker induce the system to operate in a slower way, increasing the end-to-end communication latency.

The above results prove that even when the SDN controller and the controlled switch share the same computational space, such as edge middleboxes, where RTT is expected to be very low, it remains true that the reactive mode operation unfortunately still introduces a considerable delay that can degrade the quality of IoT applications. The obtained results have also shown that the Hybrid and, particularly,

Proactive modes, with a slight increase on the configuration complexity at the network edge, can provide better performance results for all dataflow packets.

### B. *Stress Functional Tests*

We have tested our proposal using a more demanding set of testing scenarios. In this more realistic scenario, a TCP (Cubic) flow with a rate of 500 Kbit/s was originated in the host h1 and sent to the host h4. Before establishing the last communication, our proposal needs to power on the switch s2, the host h4, and enable the communications links s1-s2 and s2-h4, as resumed in Table III, 1 Flow(A): h1-h4. This table also lists the major setup steps of four more stress functional tests, with an increasingly number of simultaneous TCP (Cubic) flows.

TABLE III. STRESS FUNCTIONAL TESTS

| **1 Flow (A): h1-h4** | **1 Flow (B): h1-h5** | **2 Flows: h1-h4; h1-h5** | **3 Flows: h1-h4; h1-h5; h1-h6** | **15 Flows: 5x(h1-h4); 5x(h1-h5); 5x(h1-h6)** |
|---|---|---|---|---|
| Power on s2<br>Power on h4<br>Connect s1-s2<br>Connect s2-h4<br>Inject 1 TCP flow of 500 Kb/s | Power on h5<br>Connect s2-h5<br>Inject 1 TCP flow of 500Kb/s | Power on s2<br>Power on h4<br>Connect s1-s2<br>Connect s2-h4<br>Power on h5<br>Connect s2-h5<br>Inject 2 TCP flows of 500Kb/s each | Power on s2<br>Power on h4<br>Connect s1-s2<br>Connect s2-h4<br>Power on h5<br>Connect s2-h5<br>Power on h6<br>Connect s2-h6<br>Inject 3 TCP flows of 500Kb/s each | Power on s2<br>Power on h4<br>Connect s1-s2<br>Connect s2-h4<br>Power on h5<br>Connect s2-h5<br>Power on h6<br>Connect s2-h6<br>Inject 15 TCP flows of 100Kb/s each |

The major performance results of these stress functional are visualized in Table IV. The Delay1 represents the average time interval between the message ARP Request and the message ARP Reply of each end-to-end communication pair. The Delay2 represents the average time interval between the SYN message and the SYN-ACK message of each TCP logical connection used during the performed test.

TABLE IV. STRESS FUNCTIONAL TESTS COMPARISON (MS)

| | **1 Flow(A)** | **1 Flow(B)** | **2 Flows** | **3 Flows** | **15 Flows** |
|---|---|---|---|---|---|
| **Delay1** | 1063 | 366 | 2100 | 2072 | 2083 |
| **Delay2** | 18 | 15 | 25 | 54 | 135 |

In the scenario with a single TCP flow (test 1 Flow(A) in Table IV), the system requires 1063 ms (Delay1) to power on switch s2, connect it to the SDN controller, power on host h4, connect s1 to s2 and connect h4 to s2. When we add a second TCP flow (test 2 Flows in Table IV), the Delay1 almost doubles it value in relation to the scenario with a single TCP flow. Nevertheless, the Delay1 seems to stabilize when we compare the scenario with three or 15 TCP flows against the case with two TCP flows.

In the cases with a single TCP flow, the system has a latency of 15/18 ms (Delay2 in Table IV) due to the reactive control made by the SDN controller over the TCP flow. For the cases with two or three TCP flows, when we compare these with the scenario with a single TCP flow, the value of Delay2 increases linearly with the number of active flows. Nevertheless, when we compare the increasing trend on Delay2 between 3 Flows and 15 Flows tests, it suggests an increasing slope below the slope of the three initial tests.

From the results obtained in the set of stress functional tests, one can conclude the time to power on the switch s2 and connect it to the SDN Controller (i.e. Delay1 for 1Flow(A) in Table IV) is higher than the time to power on the host device and the communication links (i.e. Delay1 for 1Flow(B) in Table IV). In addition, our solution when controlling a gradual increase on the number of TCP flows showed a scalable and acceptable performance degradation, essentially in the highest delay component of the system (i.e. Delay1 in Table IV).

### C. *IoT Functional Tests*

We have made functional tests on the current studied SDN-based solution enhanced by the Broker application, but now in a scenario (see Figure 5) where IoT devices are communicating via MQTT-SN, a very popular IoT protocol. The reader should note that the Broker application is completely independent of the EMQ Broker of MQTT-SN.

For testing the current scenario, two Python scripts have been written to run at IoT nodes. The first node is LXC3 and behaves as a client. The second node is LXC1 and behaves like a virtualized sensor. The client script (askTemperature.py) requests a temperature value by publishing a question string "What is the temperature?” into the topic with the name askTemp. The client LXC3 then waits for the reply on a different subscribed topic, with the name sendTemp.

After the Broker executes a Python script (sendTemperature.py) that starts container LXC1, which behaves as a gateway to a real sensor. The previous script subscribes the topic askTemp, and when it receives the temperature request, fetches the temperature value from the IoT sensor, and publishes that value on the topic sendTemp, previously subscribed by the LXC3 client.

Figure 13 represents the publishing of a request message by LXC3 client and the associated response sent by LXC1 virtual sensor, after the Broker application has started it. Finally, the bottom of Figure 13 visualizes the situation after the client has received the temperature value.

```
root@control:~# python2.7 askTemperature.py
Message sent: What is the temperature?
```

```
root@sensorTemp:~# python2.7 sendTemperature.py
Message received: What is the temperature?
Message sent: 20 degrees
```

```
root@control:~# python2.7 askTemperature.py
Message sent: What is the temperature?
Message received from sendTemp: 20 degrees
```

Fig.13. MQTT-SN Interaction: i) request from the client (top), ii) request received by the sensor and its reply, iii) reply received by the client.

As a partial conclusion, the use of MQTT-SN protocol allows client/sensor interaction with a temporal decoupling. This protocol decoupling is fundamental because in a sensor network some nodes could be temporarily offline. This protocol also avoids the session data to be stored in an external device. These interesting results can be applied to other IoT protocols such as Constrained Application Protocol.

### D. *Congestion Control Tests*

It is expected that the advent of IoT devices at a large scale will increase the data volume across the global network, creating congestion issues. Considering this potential problem, we have tested the three types of congestion algorithms normally used at the Transport layer: loss-based (cubic), delay-based (vegas), and hybrid (bbr). In addition, we have added some realistic constrains to our testbed, using the traffic control tool. In this way, we have configured a generic bottleneck link between switches s1 and s2, with a maximum rate of 10Mbps. In addition, we have configured the access to the nodes behaving as servers with distinct delays: h4 (10.0.0.4) with an access delay of 10ms; h5 with an access delay of 60ms and h6 with an access delay of 110ms. Then, we have injected two TCP flows destined to each server, with a total number of six TCP flows. In each test, all the flows share the same congestion algorithm, and each flow uploads 1Mbyte of data to the respective server. The Figure 14 shows the script to test cubic performance.

```
iperf -c 10.0.0.4 -n 1M -P 2 -Z cubic &
iperf -c 10.0.0.5 -n 1M -P 2 -Z cubic &
iperf -c 10.0.0.6 -n 1M -P 2 -Z cubic &
```

Fig.14. Script commands to generate the diverse TCP (Cubic) flows.

Figure 15 shows the aggregated rate of each congestion algorithm. The congestion algorithm with the highest aggregated rate was bbr [28]. This algorithm measures both the bottleneck bandwidth and round-trip propagation time. Using this hybrid operation for detecting and controlling the network congestion, it offered the highest utilization of the available network resources despite our testing scenario having constrains in both rate and delay. In addition, the other two alternatives, i.e. cubic and vegas, showed both a period (Figure 15, samples 5-6) when their rate were very low in comparison with the bbr rate. Considering the expected high heterogeneity of IoT devices and communication protocols in future networking environments, we envisage the need of developing novel congestion algorithms well-suited to the presence of other congestion control algorithms [29].

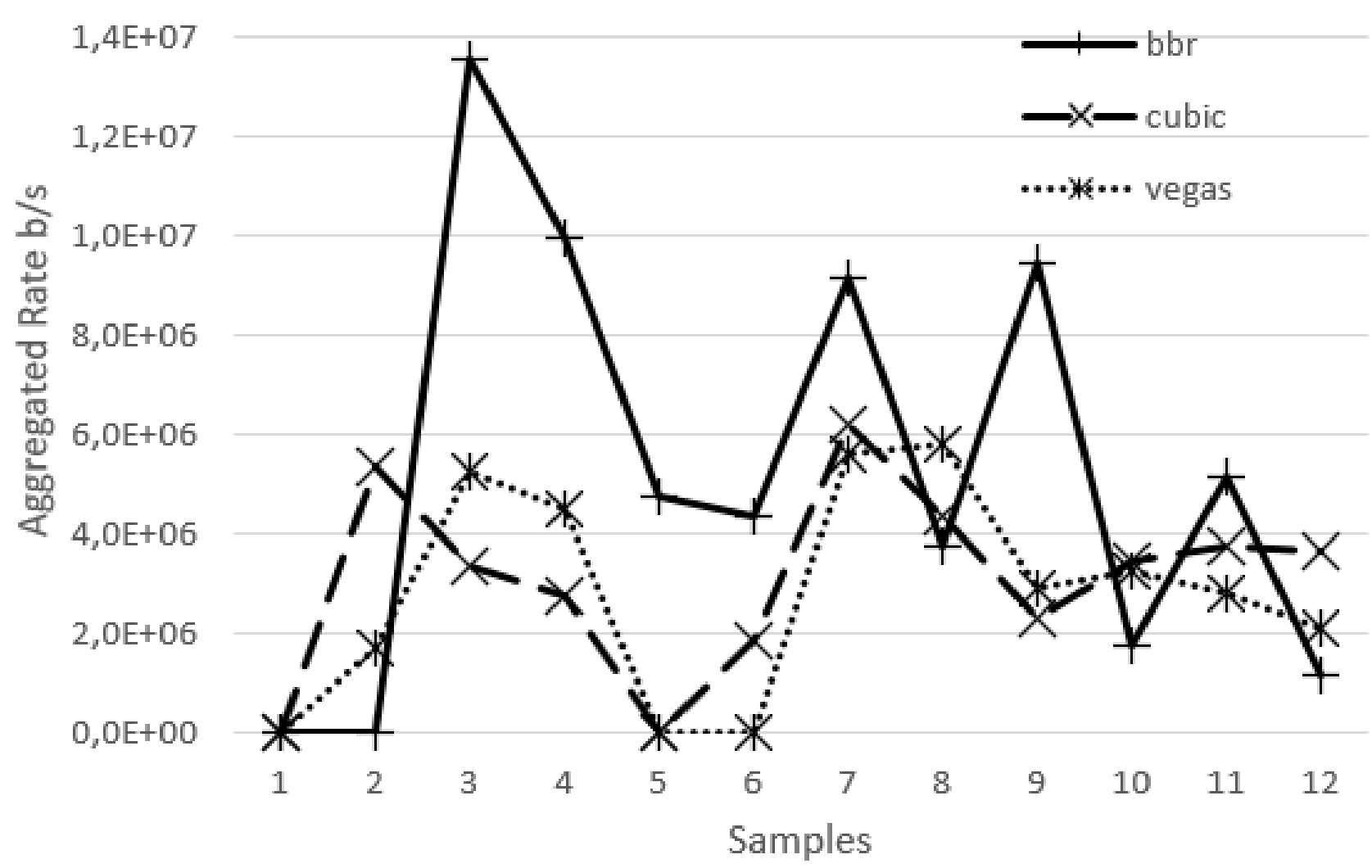

Fig.15. Aggregated rate of the diverse congestion algorithms under study.

## V. Conclusions and Future Work

The paper presents an SDN based activation policy for services provided at the edge of the network, targeting specifically IoT environments. The main idea of the paper is to push SDN controller at the edge (close to where IoT devices would be) and define a set of policies to activate dynamically data paths to enable communication flows between applications and IoT devices. The paper describes a small-scale testbed that has been implemented for testing these concepts. It also discusses the results of a set of experiments where different activation policies are used and different application-level scenarios are considered. While our testbed could be sufficient for a proof of concept, it is not sufficient for a comprehensive performance evaluation of our solution. To mitigate this limitation further work is needed.

Future trends on software-defined solutions [30] augmented by Broker agents for managing available resources on IoT networks can be namely as follows: i) extend the performance evaluation of our current testbed, using high-scalable use cases; ii) study scenarios where the SDN control plane is able to activate edge containers in various network edge domains, creating a microservice federation system.

**Funding:** The current work is funded by FCT/MCTES through national funds and when applicable co-funded EU funds under the project UIDB/50008/2020.

**Acknowledgments**: The authors acknowledge the support given by Instituto de Telecomunicações, Lisbon, Portugal.